\documentclass[aps,prl,showpacs,amssymb,twocolumn,floatfix]{revtex4}
\usepackage{graphicx}%

\def\loosen{\def\baselinestretch{1.3}\small\normalsize}
 
\begin{document}
\bibliographystyle{apsrev}
 

\loosen

\title{Numerical simulations studies of the convective instability onset in a
supercritical fluid}

\author{A. Furukawa$^1$,  H. Meyer$^2$ and A. Onuki$^1$}

\address{1. Department of Physics, Kyoto University, Kyoto 606-8502, Japan \\2.
Department of Physics, Duke University, Durham, NC 27708-0305, USA. }


\date{11/17/04}

\begin{abstract}

Numerical simulation studies in 2D with the addition of noise are reported for the 
convection of a supercritical fluid, $^3$He, in a Rayleigh-B\'{e}nard cell where the fluid
parameters and cell height $L$  are the same as in published laboratory experiments. The
noise addition  is to accelerate the instability onset after starting the heat flow across
the fluid, so as to bring simulations into better agreement with experimental observations.
Homogeneous temperature noise and  spatial lateral periodic temperature variations in the
top plate were programmed into the simulations.  A speed-up in the instability onset was
obtained, which was most effective through the spatial temperature variations with a period
of 2$L$, close to the wavelength of a pair of convections rolls. For a small amplitude of
0.5 $\mu$K, this perturbation gave a semiquantitative agreement with experimental
observations. Results for various noise amplitudes are presented and discussed in relation
to predictions by El Khouri and Carl\`{e}s.


\end{abstract}

\pacs{44.25.+f, 47.27.Te, 64.70.Fx}

\maketitle

\section{Introduction}

In recent papers, convection experiments of supercritical $^3$He  in a Rayleigh-Benard cell
with a constant heat current $q$ were reported\cite{Kogan:M:2001,Meyer:K:2002}. After  $q$
is started, the temperature drop
$\Delta T(t)$ across this highly compressible fluid layer increases from zero, an
evolution accelerated by the ``Piston Effect"
\cite{Onuki:F:1990,Zappoli:B:G:LeN:G:B:1990,Zappoli:1992}. Assuming that
$q$ is larger than a critical heat flux necessary to produce fluid instability,
$\Delta T(t)$ passes over a maximum at the time $t = t_p$, which indicates that the
fluid is  convecting and that plumes have reached the top plate. Then truncated or damped
oscillations, the latter with a period
$t_{\rm osc}$, are observed under certain conditions before steady-state conditions
for convection are reached, as described in refs.\cite{Kogan:M:2001,Meyer:K:2002}. The
scenario of the damped oscillations, and the role of the ``piston effect" has been
described in detail in refs.\cite{Furukawa:O:2002} and \cite{Amiroudine:Z:2003} and
will not be repeated here.  The height of the layer in the RB cell was $L$ = 0.106 cm
and the aspect ratio $\Gamma$=57. The $^3$He convection experiments along the critical
isochore extended  over a range of reduced temperatures between 5$\times 10^{-4} \leq
\epsilon \leq
$ 0.2, where $\epsilon  = (T - T_c)/T_c$ with $T_c$ = 3.318 K, the critical temperature. The
truncated - or damped oscillations were observed for $\epsilon \geq$ 0.009  and over
this  range the fluid compressibility varies by a factor of about 30.

The scaled representation of the  characteristic times $t_{\rm osc}$ and $t_{\rm p}$ versus
the Rayleigh number, and the comparison with the results from simulations has been
described in ref.\cite{Furukawa:M:O:K:2003}. Good agreement for
the period $t_{\rm osc}$was reported. However a systematic discrepancy for the times
$t_{\rm p}$ shows that in the simulations the development of convection 
is retarded compared to the experiments. This effect increases with decreasing 
values of
$[Ra^{\rm corr} - Ra_c]$, where
$Ra^{\rm corr}$ is the  Rayleigh number corrected for the adiabatic temperature
gradient as defined in refs.\cite{Kogan:M:2001,Meyer:K:2002} and
$Ra_c$ is the critical Rayleigh number for the experimental conditions, 1708. This is shown
in Fig.1 of ref.\cite{Furukawa:M:O:K:2003}, in particular in Fig.1b)  for
$\epsilon$ = 0.2 and $q$ = 2.16$\times 10^{-7}$ W/cm$^2$ ($[Ra^{\rm corr} - Ra_c]$ =
635), where an experimental run is compared with simulations for the same parameters.
Here clearly the profile
$\Delta T(t)$ from the simulations shows the smooth rise until the steady-state value,
$\Delta T = qL/\lambda$ = 95 $\mu$K has been reached, where $\lambda$ is the  thermal
conductivity. Only at t $\approx$ 90 s. does  convection develop, as shown by a
sudden decrease of $\Delta T(t)$. By contrast, the experimental profile shows a much
earlier development of convection. Fig.1 of ref.\cite{Furukawa:M:O:K:2003} is
representative for the observations at low values of $[Ra^{\rm corr}-Ra_c]$. At
high values, both experiment and simulations show the convection development to take place
at comparable times, as indicated in Fig.5b) of ref.\cite{Furukawa:M:O:K:2003}, and
specifically in Fig.2 a) of ref.\cite{Amiroudine:Z:2003}, where $[Ra^{\rm corr}-Ra_c]$
=4.1$\times 10^5$.  It is the purpose of this report to investigate the origin of this
discrepancy by further simulation studies.

\section{Convection onset calculations, simulations and comparison with experiments}

El Khouri and Carl\`{e}s\cite{ElKhouri:C:2002} studied theoretically the stability
limit of a supercritical fluid in a RB cell, when subjected to a heat current $q$
started at the time $t$ = 0. Their fluid was also
$^3$He at the critical density, and the same parameters as in ref.\cite{Kogan:M:2001}
were used. They calculated  the time
$t_{\rm instab}$ and also the corresponding $\Delta T(t_{\rm instab})$ for the onset
of fluid instability  and they determined the modes and the wave vectors of the
perturbations for different scenarios of $q$ and
$\epsilon$. For $t>t_{\rm instab}$ inhomogeneities in the RB cell and noise within the
fluid will produce perturbations which will grow, from which the
convection will develop. An indication of the growth of convection is a
deviation of the
$\Delta T(t)$ profile in the experiments or in the simulations from the
calculated curve for the stable fluid (see for instance Eq.3.3 of
ref\cite{Furukawa:O:2002}). It is readily seen from simulation profiles such as
Fig.1a) and b) in ref.\cite{Furukawa:O:2002} that the deviation  becomes significant for
$t$ only slightly below $t_{\rm p}$ - the maximum of $\Delta T(t)$. In simulations, the
effective start of convection can also be seen from snapshots in 2D of the fluid
temperature contour lines at various times, as shown in Fig. 5  of
ref.\cite{Chiwata:O:2001}.

P.Carl\`{e}s \cite{Carles:2003} has argued that the reason for the discrepancy for
the time $t_{\rm p}$ between experiment and simulation is that in the former, the physical
system has noise and inhomogeneities which cause the perturbations beyond 
$t_{\rm instab}$ to grow into the developed convection. By contrast simulations
have a much smaller noise. Therefore in the simulations the perturbations
take a longer time to grow than in the physical system, leading to a larger
$t_{\rm p}$ than observed. Carl\`{e}s' comment led us to try as a first step imposing a
thermal random noise on the top plate of the RB cell, which was to simulate
 fluctuations in the upper plate temperature control of the laboratory
experiment. The temperature of the plate was assumed to be uniform, because of the
large thermal diffusivity $D_T \approx 2\times 10^4$ cm$^2$/s.  of the copper plate
in the experiments.  Accordingly simulations were carried out by the numerical method
described in ref.\cite{Furukawa:O:2002} with a homogeneous time-dependent temperature random
fluctuation of given rms amplitude imposed on the upper plate. This implementation
consisted in adding or subtracting randomly temperature spikes
$T_t$ at the time
$t$ with a programmed rms amplitude at steps separated by 0.02 s. This interval is
much larger than the estimated relaxation time of the top plate over a distance $2L$,
approximately  the wavelength of convection roll pair. Values of the variance A =
$\sqrt{<(T_t -<T_t>)^2>}$  were chosen between 0 and 40
$\mu$K. The range of the A values  was taken well beyond the estimated
fluctuation rms amplitude during the experiments\cite{Kogan:M:2001} of
$\approx 1\mu K/
\sqrt{Hz}$. Three representative curves with 0, 3 and 40 $\mu$K are shown in
Fig.1a) by dashed lines for  $\epsilon$ = 0.2 for $q$= 2.16$\times10^{-7}$ W/cm$^2$,
$L$ = 0.106cm and $\Gamma$ = 5.1. For this value of $q$, the calculation by El Khouri
and Carl\`{e}s \cite{ElKhouri:C:2004} give $t_{\rm instab}$ = 6.3 s and 
$\Delta T(t_{\rm instab})$ = 75 $\mu$K. In the simulation without imposed noise, the start
of convection has therefore been considerably delayed relative to $t_{\rm instab}$. The
injection of random noise has a significant effect in developing convection at an
earlier time.  In Fig.1a) the three curves are also compared with the experimental
one, shown by a solid line. Here we have not incorporated into the simulations the delay
affecting the experimental temperature recording, so that they could be intercompared more
readily, and also with predictions\cite{ElKhouri:C:2004} However this operation will be
presented in Fig.4.  Further simulations with added random noise were carried out
for $\epsilon$ = 0.2 and 0.05 where the $\Delta T(t)$ time profiles are not shown here.

Fig.2a) shows a plot of the time of the developed convection,  represented by $t_{\rm p}$, 
versus the random rms amplitude A for three series of simulations, all taken for a cell with
$\Gamma$ = 5.1. They are a) and b) $\epsilon$ = 0.2, $q$ = 2.16 and
3.89$\times10^{-7}$W/cm$^2$, and c) $\epsilon$ = 0.05, $q$ = 60 nW/cm$^2$, ([$Ra^{\rm corr}
- Ra_c$] = 635, 1740 and 4200). The simulation results, shown by solid circles, are
compared with the experimentally observed $t_{\rm p}$  shown by horizontally dot-dashed
lines. It can be clearly seen that noise imposition, which creates a vertical disturbance
across the fluid layer, reduces the time of convection development. While the decrease of
$t_{\rm p}$ is strong for small values of A, it saturates at a certain  level of noise
amplitude. The gap between simulations and experiment increases with a decrease of
[$Ra^{\rm corr} - Ra_c$], namely as the fluid stability point is approached. A ``critical
slowing down" is seen in the effectiveness of the  perturbations in triggering the
instability. Hence this mode of noise introduction fails, because its amplitude is limited
to the vertical z direction and it evidently couples only weakly into the convective motion.

\begin{figure}
\includegraphics[width=3.2in]{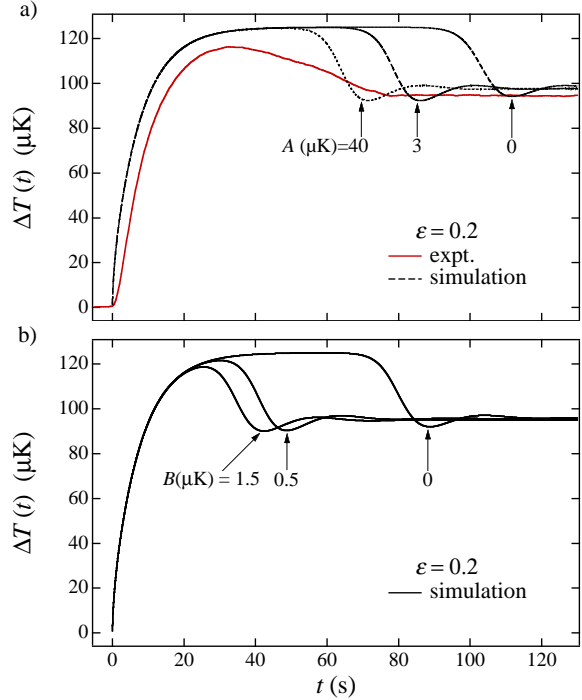}
\caption{a) the temperature profile $\Delta T(t)$ from experiments (solid line with
noise) and from several simulations (dashed lines) at $\epsilon$ = 0.2, $q$=
2.16$\times 10^{-7}$ W/cm$^2$. In the simulations, $\Gamma$=5 and uniform temperature
noise has been imposed on the top plate with  variance $A(\mu K)$ = 0, 3 and 40, as
described in the text. b) Temperature profile $\Delta T(t)$ from  several simulations
at $\epsilon$ = 0.2,
$q$= 2.16$\times 10^{-7}$ W/cm$^2$, $\Gamma$ = 8 and imposed lateral periodic, time
independent  temperature variations on the top plate with  period $2L$ and amplitude
$B  (\mu K)$ = 0, 0.5 and 1.5.}
\end{figure}

\begin{figure}
\includegraphics[width=3.2in]{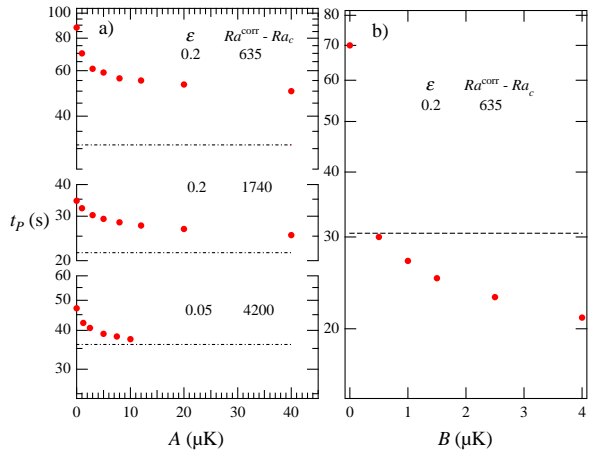}
\caption{a) The time for effective development of convection, characterized by $t_{\rm p}$,
versus $A$ (homogeneous temperature noise imposed on the top plate). The horizontal
dot-dashed lines indicate the observed  $t_{\rm p}$, corrected for instrumental recording
delay. b) The time for effective development of the convection, labeled by
$t_{\rm p}$ versus $B$ (lateral time-independent periodic temperature variations). The
horizontal dashed line indicates
$t_{\rm p}$ as obtained by experiment, corrected as before.}
\end{figure}

In parallel with the present experiments, S. Amiroudine\cite{Amiroudine:2004}  also
carried out a systematic study of simulations on supercritical $^3$He in a RB cell
for several values of $\epsilon$ and $q$. He used a numerical scheme based on the exact
Navier Stokes equation as described in ref.\cite{Amiroudine:Z:2003}. The resulting profiles
$\Delta T(t)$ could be compared with those from experiments done under nearly the same
conditions. In his simulations, homogeneous temperature random noise was again imposed on
the top plate. The shift in $t_{\rm p}$  showed less systematic trends than in the results
described in this report. However for the same values of
$\epsilon$ and $q$ as those reported above, and at zero noise, the $t_{\rm p}$ values
tended to be somewhat smaller than in the results of Fig 2a).

Here we mention that the onset of convection in the simulations is further
influenced by the aspect ratio $\Gamma$. The simulations described above, but without
noise, were carried out in a cell $\Gamma$ = 5.1 having periodic lateral
boundaries. Further simulations with zero noise for $\epsilon$ = 0.2  with
$\Gamma$ =  8.0, 10.2, 20.5 and 41.0  were carried out, and showed a decrease of the
 convection development time from
$\approx$ 90 s,   tending to a constant value of $\approx$ 60 s. above
$\Gamma$ = 20. This shift in the onset of instability is due to the decreased
finite size effect which the rising plumes experience with increasing
$\Gamma$, in spite of the periodic boundary conditions. This can be seen by comparing the
curves labeled ``O" in Figs 1a and 1b with $\Gamma$ = 5 and 8 respectively.

The next step in our attempts, stimulated by communications with P. Carl\`{e}s, was
introducing perturbations into the simulations via some time-independent lateral
variation proportional to sin (2$\pi x/P$) where
$P$ is the period. We opted to introduce again a
temperature variation in the top plate with an  amplitude B (in $\mu$K) and period
$P$=2$L$, nearly the same as the wavelength of  a pair of convection rolls. The
temperature of the bottom plate was kept homogeneous.
 This ``Gedanken Experiment" implies that the material of the top plate permitted a
temperature inhomogeneity, which of course is not realized in the experiment. However a
small lateral temperature excursion can trigger the same kind of non-homogeneous
perturbations as those which, in the real experiment, provoke the onset of convection. One
possible origin of such perturbations, besides thermal noise, could be the roughness of the
plates or their slight deviation from parallelism. Such geometrical defects could of course
not be implemented in the numerical simulations with the meshsize used, which is why we
elected to force a small temperature perturbation instead, with similar effects on the
onset. As a control experiment, we also made a simulation with
$P=L$.

Fig 1b) shows representative profiles $\Delta T(t)$  for the parameters $\epsilon$ =
0.2 and $q$= 2.16$\times 10^{-7}$ W/cm$^2$ and with B = 0, 0.5 and 1.5 $\mu$K, and for
$\Gamma$=8. As B is increased from
zero, there is a large decrease in the time for convection development, represented
by $t_{\rm p}$, which  is plotted versus B in Fig 2b). The horizontal dashed line shows the
$t_{\rm p}$ from the experiment, and this plot is to be compared with Fig. 2a). For an
inhomogeneity amplitude of only B= 0.5$\mu$K, $t_{\rm p}$ is nearly the same for simulations
and experiment. By contrast, simulations with B=2$\mu$K and 
$P$=$L$ (not presented here) show no difference from those with B=0. Hence the
nucleation of the convection is accelerated if the period is in approximate resonance
with the wavelength of a convection roll pair. The values of steady-state $\Delta T$
and
$t_{\rm osc}$ are only marginally affected by the noise.

We note  from  Fig.1b) that the simulation curve  calculated for B = 0 shows
 the fluid  not convecting until $\approx$ 70 s. For the curves with B= 0.5
$\mu$K., the start of deviations from the stable fluid curve cannot be estimated well
from Fig.1b) but is readily obtained from the data files, which tabulate $\Delta T(t)$
to within 1 nK. For B = 0.5 $\mu$K, systematic deviations $\delta\Delta T(t,B)\equiv[\Delta
T(t$,B=0) - $\Delta T(t$,B)] increase rapidly from 1 nK for $t > $8 s (where $\Delta
T\approx$85 $\mu$K), a value comparable with the predicted $t_{\rm instab}$= 6.3 s., 
$\Delta T(t_{\rm instab})$ = 75$\mu$K\cite{ElKhouri:C:2004}. However a comparison with
predictions becomes more uncertain as B is increased and no longer negligible
compared with the steady-state $\Delta T$. Then it is expected that the base Piston-Effect
heat flow will become itself influenced by the perturbations. In that case the stability
analysis in \cite{ElKhouri:C:2004} becomes irrelevant, since the base flow, the stability
of which is analysed, has been significantly modified by the perturbations. We also note
that the time interval $\delta t \equiv [t_{\rm p}-t_{\rm instab}$] between the first sign
of instability ($\delta\Delta T > 0 $) and
$t_{\rm p}$ is $\approx$ 20 s, and  roughly independent of B. This represents approximately
the period taken by the convection to develop and for the plumes to reach the top plate
boundary.

In Fig.3, we present a series of 2D ``snapshots"  at various times for the simulation
with B = 0.5$\mu$K,  showing the temperature contour lines (in color) for the RB
cell. The ``warm" side is shown by red, $T(t, z=0)$ and the ``cold"
side by mauve, $T(z=L)$ =  const. At $t$= 8 s. the fluid  instability  has just
started near the top of the layer, while near the bottom the
\begin{figure}
\includegraphics[width=3.0in]{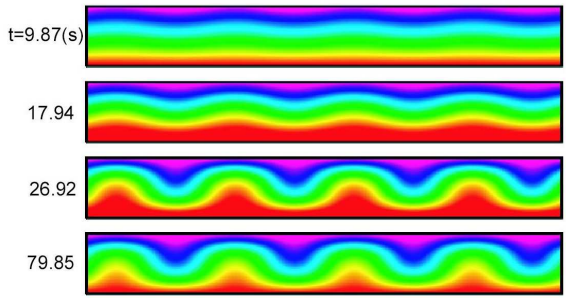}
\caption{Snapshots in 2D for the RB cell with an aspect ratio of 8, of simulations
with B = 0.5 $\mu K$ at various times $t$ after starting the heat current $q$. The
temperature contour lines and their evolution  are described in the text. At the time
$t$ = 9.9s, the fluid instability has just started near the top.} 
\end{figure}
temperature contour lines are still horizontal. At t =  27 s., where the
peak of $\Delta T(t)$  at $z=L$ has been reached, the warm plumes have reached the top
plate, and the ``cold" piston effect is about to start, causing  the bulk fluid temperature
 to drop and  $\Delta T(t)$  to decrease. The transient process  continues
with  damped oscillations of $\Delta T(t)$. Steady state convection is reached at t=
80s, with a pair of convection rolls having a wavelength of $\approx$ 2$L$, as expected.

In Fig.4 we show the profiles $\Delta T(t)$ from the experiment and from the
simulations with a periodic perturbation amplitude B = 0.5 $\mu$K. For an optimal
comparison, the delay affecting the experimental  temperature recording was
incorporated into the simulation curve. For this, the delay function with the
instrumental time constant $\tau$ = 1.3 s. \cite{Kogan:M:2001} was folded into the
simulation curve by a convolution method. This operation  retards the initial rise of
the temperature drop by the order of  2-3 seconds, and brings both experiment and
simulations into fair agreement in the regime where the fluid is stable. The time $t_{\rm
p}$ for the maximum is now closely the same for both experiments and simulations. However
beyond the predicted instability time $t_{\rm instab}$= 6.3 s., the
experimental curve starts to deviate more rapidly with time than do the numerical
simulations from the calculated curve for the fluid in the stable regime. As discussed
previously\cite{Meyer:K:2002}, for these parameters of $\epsilon$ and $q$ the experiment
does not show damped oscillations, which are observed for higher values of $q$. In the
steady-state, the agreement is very good.

\begin{figure}
\includegraphics[width=3.0in]{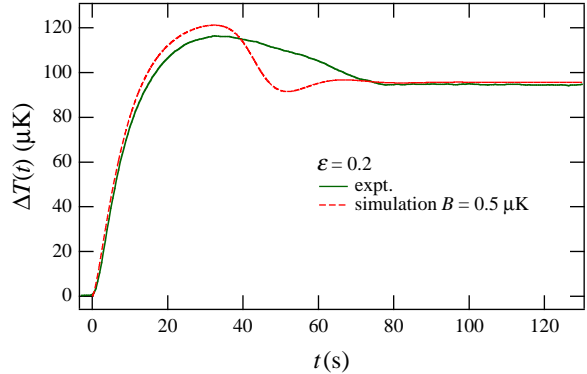}
\caption{Comparison of the profile $\Delta T(t)$ from experiment and from simulations
with  B = 0.5 $\mu K$ . To make the comparison realistic, the simulations have been
convoluted with the same ``instrumental" delay time $\tau$ =1.3 s. which has
influenced the shape of the experimental curve.} 
\end{figure}

Our goal has been to show that  injecting a small temperature perturbation into the
top plate, produces for the simulations an earlier start in the convective
instability, which becomes consistent with experimental observations. For this, we
have limited ourselves to an example at a low value of
$[Ra^{\rm corr} - Ra_c]$, where the delay has been  particularly large with respect to
the experiment. 

\section{Summary and Conclusion}
 We have presented a comparison of numerical simulations with experimental data
investigating the transient to steady convection after the start of a heat current
through a supercritical $^3$He layer in a RB cell. Here the temperature drop $\Delta
T(t)$ across the fluid layer versus time $t$ was studied. The aim was to
understand and to reduce the discrepancy between experiment and simulations in the
time of the convection development, as detected by 
$\Delta T(t)$. Simulations
for one set of fluid parameters (where the largest discrepancy had been observed) are
reported with imposed temperature variations on the top plate. Satisfactory results were
obtained for  spatial lateral temperature variations with an  amplitude of 0.5 $\mu$K and a
period approximately equal to that of the wavelength of a convection roll pair. As the
perturbation amplitude is further increased, the development of convection occurs earlier
than the observed one.

\section{Acknowledgment} The authors are greatly indebted to P. Carl\`{e}s for
stimulating correspondence and suggestions, to F. Zhong for help with figures
formatting and the convolution program in Fig.3 and to R.P. Behringer and P. Carles for
useful comments on the manuscript.  The interaction with S. Amiroudine, who conducted
numerical simulation in parallel with present investigations is greatly appreciated. The
work is supported by the NASA grant NAG3-1838 and by the Japan Space Forum H12-264.




\begin{thebibliography}{99}

\bibitem{Kogan:M:2001} A.B. Kogan and H. Meyer, Phys.\ Rev.\ E {\bf 63}, 056310
(2001).

\bibitem{Meyer:K:2002} H. Meyer and A.B. Kogan,  Phys.\ Rev.\ E {\bf 66},056310
(2002).

\bibitem{Onuki:F:1990} A. Onuki and R.A. Ferrell, Physica A {\bf 164},  245 (1990).

\bibitem{Zappoli:B:G:LeN:G:B:1990}B. Zappoli, D. Bailly, Y. Garrabos,  B. le Neindre,
P. Guenoun and D. Beysens, Phys.\ Rev.\ A {\bf 41}, 2264 (1990).

\bibitem{Zappoli:1992} B. Zappoli, Phys. of Fluids {\bf 4}, 1040 (1992), B. Zappoli
and P. Carles, Eur. J. Mech. B/Fluids {\bf 14}. 41, (1995) 



\bibitem{Furukawa:O:2002} A. Furukawa and A. Onuki Phys.\ Rev.\ E {\bf 66}, 016302
(2002).


\bibitem{Amiroudine:Z:2003} S. Amiroudine and B. Zappoli, Phys.\ Rev.\ Lett.\  {\bf
90}, 105303 (2003).

\bibitem{Furukawa:M:O:K:2003} A. Furukawa, H. Meyer, A. Onuki and A.B. Kogan, Phys.\
Rev.\ E {\bf 68}, 056309 (2003)]

\bibitem{ElKhouri:C:2002} L. El Khouri and P. Carl\`{e}s,  Phys.\ Rev.\ E {\bf 66},
 066309 (2002).

\bibitem{Chiwata:O:2001} Y. Chiwata and A. Onuki, Phys.\ Rev.\ Lett.\  {\bf 87},
144301 (2001).

\bibitem{Carles:2003} P. Carl\`{e}s, private communication.


\bibitem{ElKhouri:C:2004}L. El Khouri and P. Carl\`{e}s, Private communication.


\bibitem{Amiroudine:2004} S. Amiroudine, private communication.


\end{thebibliography}
\end{document}